\documentclass[%
 aip,
 amsmath,amssymb,
 reprint,%
]{revtex4-1}

\usepackage{graphicx}
\usepackage{dcolumn}
\usepackage{bm}

\usepackage[utf8]{inputenc}
\usepackage[T1]{fontenc}
\usepackage{mathptmx}
\usepackage{enumitem}
\usepackage{textcomp}
\usepackage{color}

\begin{document}


\title{Phase Transitions in Germanium Telluride Nanoparticle Phase-Change Materials Studied by Time-Resolved X-Ray Diffraction} 

\author{Ann-Katrin U. Michel}
\affiliation{Optical Materials Engineering Laboratory, Department of Mechanical and Process Engineering, ETH Zurich, 8092 Zurich, Switzerland}%

\author{Felix Donat}
\affiliation{Laboratory of Energy Science and Engineering, Department of Mechanical and Process Engineering, ETH Zurich, 8092 Zurich, Switzerland}%

\author{Aurelia Siegfried}
\affiliation{Optical Materials Engineering Laboratory, Department of Mechanical and Process Engineering, ETH Zurich, 8092 Zurich, Switzerland}%

\author{Olesya Yarema}
\affiliation{Institute for Electronics, Department of Information Technology and Electrical Engineering, ETH Zurich, 8092 Zurich, Switzerland}%

\author{Hanbing Fang}
\affiliation{Optical Materials Engineering Laboratory, Department of Mechanical and Process Engineering, ETH Zurich, 8092 Zurich, Switzerland}%

\author{Maksym Yarema}
\affiliation{Institute for Electronics, Department of Information Technology and Electrical Engineering, ETH Zurich, 8092 Zurich, Switzerland}%

\author{Vanessa Wood}
\affiliation{Institute for Electronics, Department of Information Technology and Electrical Engineering, ETH Zurich, 8092 Zurich, Switzerland}%

\author{Christoph R. Müller}
\affiliation{Laboratory of Energy Science and Engineering, Department of Mechanical and Process Engineering, ETH Zurich, 8092 Zurich, Switzerland}%

\author{David J. Norris}
\email{dnorris@ethz.ch}
\affiliation{Optical Materials Engineering Laboratory, Department of Mechanical and Process Engineering, ETH Zurich, 8092 Zurich, Switzerland}%


\begin{abstract}
Germanium telluride (GeTe), a phase-change material, is known to exhibit four different structural phases: three at room temperature (one amorphous and two crystalline, $\alpha$ and $\gamma$) and one at high temperature (crystalline $\beta$). Because transitions between the amorphous and crystalline phases lead to significant changes in material properties (e.g., refractive index and resistivity), GeTe has been investigated as a phase-change material for photonics, thermoelectrics, ferroelectrics, and spintronics. Consequently, the temperature-dependent phase transitions in GeTe have been studied for bulk and thin-film GeTe, both fabricated by sputtering. Colloidal synthesis of nanoparticles offers a more flexible fabrication approach for amorphous and crystalline GeTe. These nanoparticles are known to exhibit size-dependent properties, such as an increased crystallization temperature for the amorphous-to-$\alpha$ transition in sub-10\,nm GeTe particles. The $\alpha$-to-$\beta$ phase transition is also expected to vary with size, but this effect has not yet been investigated for GeTe. Here, we report time-resolved X-ray diffraction of GeTe nanoparticles with different diameters and from different synthetic protocols. We observe a non-volatile amorphous-to-$\alpha$ transition between 210$^{\circ}$C and 240$^{\circ}$C and a volatile $\alpha$-to-$\beta$ transition between 370$^{\circ}$C and 420$^{\circ}$C. The latter transition was reversible and repeatable. While the transition temperatures are shifted relative to the values known for bulk GeTe, the nanoparticle-based samples still exhibit the same structural phases reported for sputtered GeTe. Thus, colloidal GeTe maintains the same general phase behavior as bulk GeTe while allowing for more flexible and accessible fabrication. Therefore, nanoparticle-based GeTe films show great potential for applications, such as in active photonics.
\end{abstract}

\maketitle

\section{\label{sec:level1intro}Introduction}

Germanium telluride (GeTe) is a metal chalcogenide that exhibits three structural phases at room temperature and one phase at high temperature.\cite{Schlieper1999,Bletskan2005} The room-temperature phases are amorphous GeTe and crystalline rhombohedrally distorted $\alpha$- as well as orthorhombic $\gamma$-GeTe. The high-temperature crystalline cubic phase is known as $\beta$-GeTe. While the atoms of such a phase-change material are covalently bonded in its amorphous (A) state, metavalent bonding can be found in the crystalline (C) states.\cite{Raty2019} These two types of bonds lead to very different properties. The amorphous phase has a relatively low optical reflectivity $R_A$ and high electrical resistivity $\rho_A$. Upon crystallization, the resistivity decreases by five orders of magnitude with $\rho_A$\,$\approx$\,10$^2$\,$\Omega$cm and $\rho_C$\,$\approx$\,10$^{-3}$\,$\Omega$cm.\cite{Jost2015} Simultaneously, the reflectance contrast $\Delta R$ defined by $[(R_C$\,-\,$R_A)$\,/\,$R_C]$\,$\times$\,100 is about 43\% in the near-infrared spectral range (wavelengths near 1\,\textmu m).\footnote{The reflectances $R_A$ and $R_C$ were calculated via Fresnel coefficients for a 100\,nm GeTe film between a silicon and an air half-space, under normal incidence, and with the dielectric function of GeTe according to Ref.~\onlinecite{Shportko2009}. The definition of the reflectance contrast $\Delta R$ was taken from Ref.~\onlinecite{Schlich2015}.}

The room-temperature amorphous phase of GeTe crystallizes at $T_{C,1}$\,=\,185$^{\circ}$C.\cite{Jost2015} Thermal annealing, optical pulses, or electrical pulses can be used to induce this structural relaxation. However, only laser or electrical pulses allow for quenching of the GeTe melt and thus, re-amorphization. The melting temperature of GeTe is $T_M$\,=\,723\,$^{\circ}$C.\cite{Schlieper1996} 

The strong property contrast between the non-volatile phases can be exploited for optical data storage and memristive memories.\cite{Raoux2010,Wright2012} The latter are one of the most promising candidates for neuromorphic computing.\cite{Boybat2018} While ternary and quaternary phase-change materials, such as germanium antimony telluride or silver indium antimony telluride, have been applied in phase-change memories, germanium telluride has gained interest for active photonics.\cite{Wuttig2017,Carrillo2019,Hail2019,Michel2020b} Furthermore, GeTe exhibits ferromagnetism with a Curie temperature of about 920$^{\circ}$C, while doping GeTe with Mn, Fe, or Cr leads to a Curie temperature $\leq$\,420$^{\circ}$C.\cite{Kriener2016} Moreover, GeTe has recently been identified as a Rashba ferroelectric.\cite{Slawinska2020}

All of the aforementioned applications and effects have been investigated either for epitaxially grown or sputtered GeTe films ranging from several tens of nanometers to several micrometers in thickness. Recently, spatially confined phase-change materials have been studied, mainly due to two opportunities. First, nanowires and nanoparticles offer an alternative approach to fabricate films of phase-change material or patterned arrays.\cite{Milliron2007,Agarwal2009,Caldwell2010,Yarema2018} Thereby, the purchase of dedicated expensive equipment (e.g. magnetron sputtering tool) can be avoided. In addition, preformed, high-aspect-ratio voids or patterns can be filled. Second, nanoscale phase-change materials allow for studying size-dependent properties of these compounds. For example, localized surface plasmon resonances have been reported for crystalline GeTe nanoparticles,\cite{Polking2013} and a bandgap increase has been observed.\cite{Michel2020,Yarema2020} Furthermore, several studies on the size-dependent shift of the crystallization temperature $T_{C,1}$ have been published, as shown for GeTe in Tab.~\ref{tab:temp}. The listed values for $T_{C,1}$ refer to the transition from the amorphous phase to the rhombohedrally distorted $\alpha$-GeTe and reveal that $T_{C,1}$ increases with decreasing particle diameter $d$. We also note that the observed crystallization temperature depends not only on the material dimensions but also on the characterization technique. For example, the drop in resistivity associated with crystallization does not require a phase change of the entire GeTe volume; a conductive crystalline channel in the film is sufficient. Another important factor is the applied heating rate $\vartheta$ throughout the measurement. A well-known example is the shifted peak temperature to higher $T$ upon increase of $\vartheta$ in differential scanning calorimetry (DSC). All of the aforementioned effects have to be taken into account when comparing the values for $T_{C,1}$ and making conclusions about size-dependent effects.

While $T_{C,1}$ has been reported for spatially confined GeTe, the high-temperature crystalline $\beta$ phase has neither been observed for ultra-small, nor initially amorphous nanoparticles so far (\textit{cf.} Tab.~\ref{tab:temp}). Here we study the reversible crystalline-to-crystalline phase transition from $\alpha$- to $\beta$-GeTe at $T_{C,2}$ and back to the $\alpha$-phase for sub-10\,nm GeTe nanoparticles, which were initially amorphous after synthesis. This is realized by collecting X-ray diffraction (XRD) patterns for repeated heating and cooling cycles of drop-casted particles, which are synthesized in their amorphous phase.

\begin{table}
\caption{\label{tab:temp}Comparison of crystallization temperatures $T_{C,1}$ and $T_{C,2}$ for different GeTe samples, determined by different characterization methods (XRD - X-ray diffraction, at a synchrotron ($_{\textnormal{s}}$) if applicable, $\rho(T)$ - resistivity measurement during heating and cooling, and DSC - differential scanning calorimetry) with varied heating rates given in $^{\circ}$C/min. The samples have either been synthesized (approx. spherical particle diameter $d$) or sputtered (approx. film thickness $t$). Initially, the GeTe has either been in its amorphous ($_A$) or crystalline ($_C$) state. The sputtered thin films are given for reference and separated by a horizontal line. $^{\dagger}$ and $^{\ddagger}$ mark a surface-oxidized and TaN-capped GeTe film, respectively.}
\begin{ruledtabular}
\begin{tabular}{llllll}
Size\,[nm] & T$_{C,1}$\,[$^{\circ}$C] & T$_{C,2}$\,[$^{\circ}$C] & Method & $\vartheta$\,[$^{\circ}$C/min]  & Ref.\\
\hline
$d_A$\,=\,1.8 & 400 & - & \textit{in-situ} XRD$_{\textnormal{s}}$ & 60 & \onlinecite{Caldwell2010}\\ 
$d_A$\,=\,2.6 & 350 & - & \textit{in-situ} XRD$_{\textnormal{s}}$ & 60 & \onlinecite{Caldwell2010}\\ 
$d_A$\,=\,3.4 & 320 & - & \textit{in-situ} XRD$_{\textnormal{s}}$ & 60 & \onlinecite{Caldwell2010}\\ 
$d_A$\,=\,3.5 & 340 & - & $\rho(T)$ & 300\,-\,1.800 & \onlinecite{Caldwell2010}\\
$d_A$\,=\,6.0 & 227   & - & \textit{in-situ} XRD & 7 &\onlinecite{Yarema2018}\\ 
                                & 170 & - & $\rho(T)$ & - & \onlinecite{Yarema2018}\\
                                & 223\,-\,240 & - & DSC & 2.5\,-\,30 & \onlinecite{Yarema2018}\\
$d_A$\,=\,8.7 & 237 & - & DSC & 5 &\onlinecite{Arachchige2011}\\ 
$d_A$\,=\,10.6 & 224 & - & DSC & 5 &\onlinecite{Arachchige2011}\\ 
$d_A$\,=\,18.5 & 209 & - & DSC & 5 &\onlinecite{Arachchige2011}\\ 
$d_C$\,=\,17.0 & - & 355 & \textit{in-situ} XRD$_{\textnormal{s}}$ & 60 &\onlinecite{Polking2011}\\ 
$d_C$\,=\,100 & - & 360 & \textit{in-situ} XRD$_{\textnormal{s}}$ & 60 &\onlinecite{Polking2011}\\
$d_C$\,=\,500 & - & 370 & \textit{in-situ} XRD$_{\textnormal{s}}$ & 60 &\onlinecite{Polking2011}\\  
\hline
$t_A$\,=\,50 & 170 & 350 & \textit{in-situ} XRD$_{\textnormal{s}}$ & 180 & \onlinecite{Raoux2009}\\
                        & 175 & - & $\rho(T)$ & 60 & \onlinecite{Raoux2009}\\                   
$t_A$\,=\,80 & 185  & - & $\rho(T)$ & 5 &\onlinecite{Jost2015}\\
$t_A$\,=\,100$^{\dagger}$ & 180  & - & $\rho(T)$ & 10 &\onlinecite{Kolb2019}\\
$t_A$\,=\,100$^{\ddagger}$ & 230  & - & $\rho(T)$ & 10 &\onlinecite{Kolb2019}\\
$t_A$\,=\,150 & 180 & - & $\rho(T)$ & 10 & \onlinecite{Mantovan2017}\\
\end{tabular}
\end{ruledtabular}
\end{table}

\section{\label{sec:level1exper}Experimental methods}
We prepared colloidal dispersions of amorphous monodisperse GeTe nanoparticles following two different protocols. All studied spherical nanoparticles had a diameter $d$\,<\,10\,nm since size-dependent crystallization had previously been identified for this size regime (Tab.~\ref{tab:temp}). 

\subsection{\label{sec:level2Synth}Nanoparticle synthesis}
The syntheses of the amorphous (A-)GeTe nanoparticles followed protocols adapted from Caldwell \textit{et al}.\cite{Caldwell2010} and reported by Yarema \textit{et al}.\cite{Yarema2018} 

\begin{figure}
\includegraphics{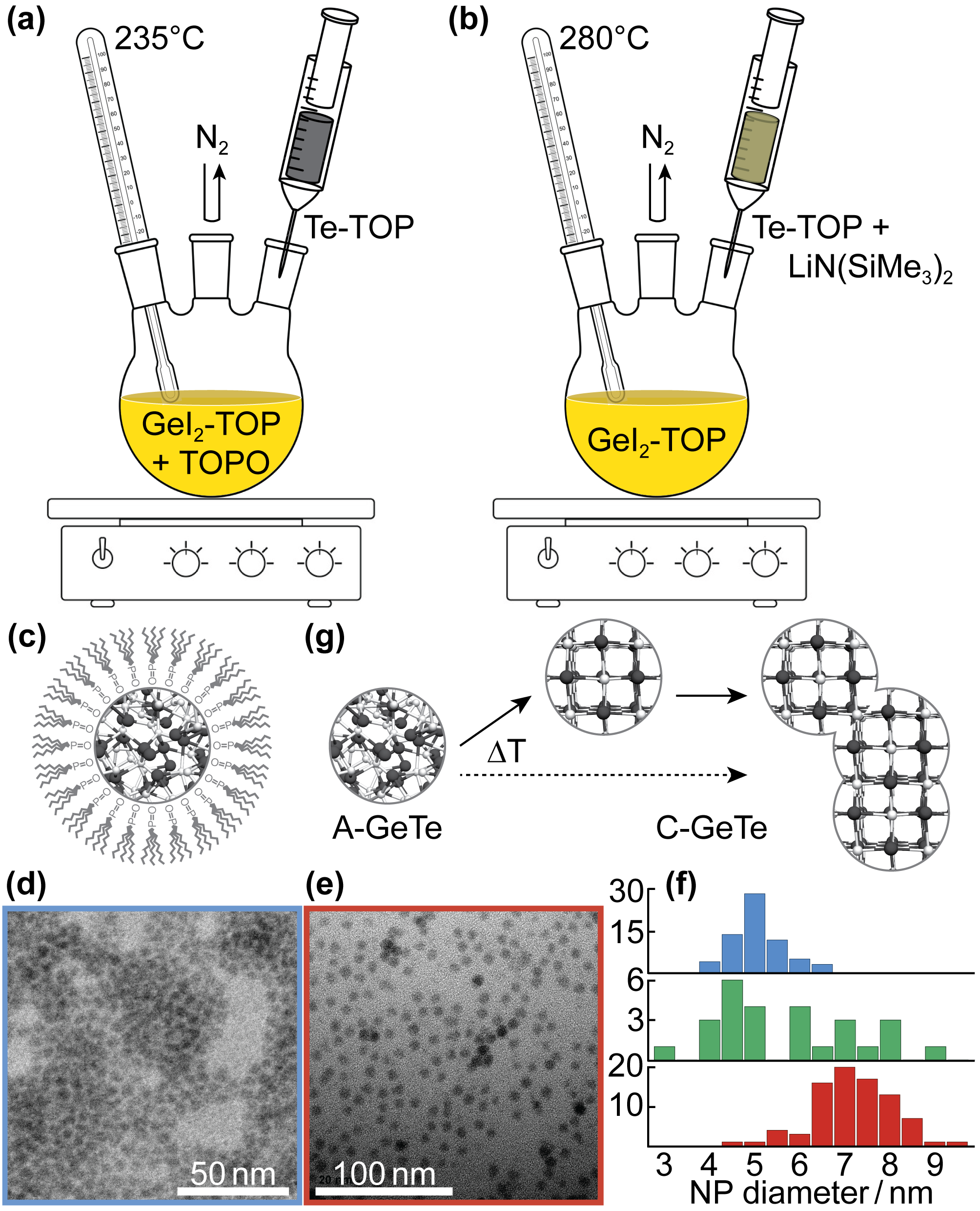}
\caption{\label{fig:synthesis} Sketch of the hot-injection methods 1, (a), and 2, (b), used for the synthesis of amorphous (A) sub-10\,nm GeTe nanoparticles, covered by organic ligands, such as trioctylphosphine (TOP), shown in (c). TEM images of the particles with an average diameter of 4.8\,nm, (d), and 6.9\,nm, (e), allow for the determination of the particle size distributions shown in (f). The top (blue) and bottom (red) distribution relate to particles from synthesis 2, (b); the middle (green) distribution is for synthesis 1, (a). (g) Heating of the A-GeTe particles with $\Delta T$\,$\geq$\,$T_{C,1}$ results in crystalline (C) GeTe particles which are likely to coalesce prior or throughout the crystallization process (dashed and solid arrow, respectively).}
\end{figure}

The first batch was synthesized through a hot-injection route as schematically shown in Fig.~\ref{fig:synthesis}(a). Anhydrous germanium(II) iodide (GeI$_2$, 163\,mg) was dissolved in 2\,ml trioctylphosphine (TOP) in a glove box and stirred overnight. The following day, 2\,g of trioctylphosphine oxide (TOPO) were added and the yellow solution was transferred to a reaction flask which was purged with nitrogen beforehand. After heating the solution to 235$^{\circ}$C, 240\,\textmu l dodecanethiol and 30\,s later 667\,\textmu l 0.75\,M Te-TOP solution (previously prepared) were injected. About 40\,s later, the color of the solution in the flask changed from yellow to dark brown, indicating nucleation of nanoparticles. After 4\,min, the reaction was terminated and the flask was cooled rapidly by acetone mist and, later, with pressurized air. The crude solution of GeTe nanoparticles was transferred air-free to the glove box where anhydrous ethanol was added (3:1). The black precipitate was separated by centrifugation (4000\,rpm, 10\,min) and dispersed in 1\,ml anhydrous chloroform. After centrifuging (4000\,rpm, 10\,min), ethanol was added to the dark brown solution (2:1). Another centrifugation step resulted in a clear liquid and a black precipitate. The latter was dispersed in 1\,ml toluene, forming a colloid that remained stable for multiple weeks. We refer to this synthetic protocol below as \textit{synthesis 1}.

The alternative synthetic approach, \textit{synthesis 2}, led to several batches with different GeTe particle sizes. This amide-promoted synthesis is schematically shown in Fig.~\ref{fig:synthesis}(b) and described in detail in Ref.~\onlinecite{Yarema2018}. While the A-GeTe nanoparticles obtained from synthesis 1 were covered by TOP ligands [\textit{cf.} Fig.~\ref{fig:synthesis}(c)], the particles from synthesis 2 were covered with an oleate shell.

From transmission electron microscopy (TEM) observations, the average size of each particle synthesis was estimated. The A-GeTe particle size available from synthesis 1 was 5.5\,$\pm$\,1.6\,nm; synthesis 2 provided A-GeTe particles with diameters 4.8\,$\pm$\,0.6\,nm, Fig.~\ref{fig:synthesis}(d), and 6.9\,$\pm$\,0.9\,nm, Fig.~\ref{fig:synthesis}(e). It has to be noted that synthesis 2 led to particles with a much narrower size distribution, as visible in Fig.~\ref{fig:synthesis}(f): the green size distribution refers to the particles from synthesis 1 and the blue and red size distributions refer to the particles from synthesis 2.

Upon annealing, the A-GeTe particles will relax into the crystalline phase if $\Delta T\,>\,T_{C,1}$. However, due to the large surface-to-volume ratio of small nanoparticles, coalescence is energetically favorable. Thus, coalescence of sub-10\,nm particles has been reported either at temperatures above $T_{C,1}$,\cite{Yarema2018} or at lower temperatures. Thus, it occurs prior or throughout crystallization [\textit{cf.} Fig.~\ref{fig:synthesis}(g)].\cite{Keitel2016}

\subsection{\label{sec:level2XRD}Time-dependent X-ray diffraction}
Samples were prepared for each particle size by repeated drop casting onto a circular quartz substrate (diameter 1.3\,cm). The sample thickness and weight, including the nanoparticles, ligands, and residual solvent, were not determined. However, the sample deposition was conducted with a particle concentration of about 5\,mg/ml and an estimated deposited amount of 5\,mg. To avoid oxidation, the colloid was deposited air-free in a nitrogen glove box. The samples were mounted in an Anton Paar XRK 900 reactor chamber that was purged with nitrogen (flow rate: 200\,ml/min, measured at ambient temperature and pressure) throughout the entire measurement. The samples were characterized with a PANalytical Empyrean diffractometer equipped with a X’Celerator Scientific ultrafast line detector and Bragg-Brentano HD incident beam optics. The instrument was operated at 45\,kV and 40\,mA using Cu K$\alpha$ radiation (1.54060\,\AA). The temperature was measured and controlled in the vicinity of the sample using a type K thermocouple; separate control measurements with a second thermocouple placed at the exact position of the sample indicated that the temperature difference between the two thermocouples was <\,5$^{\circ}$C for temperatures $T$\,<\,800$^{\circ}$C.

\begin{figure}
\includegraphics{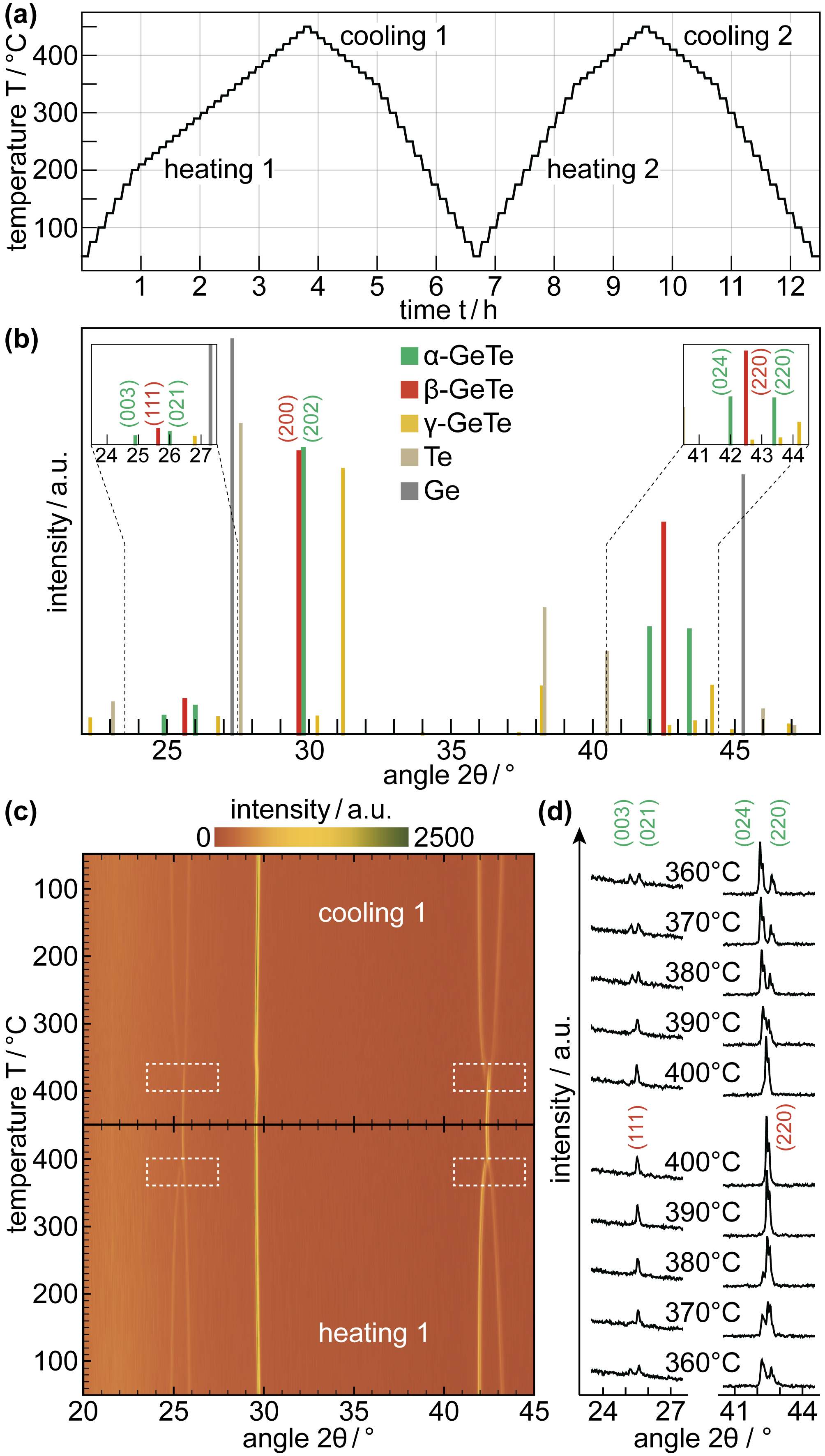}
\caption{\label{fig:tempcurve} (a) XRD chamber temperature $T$ as a function of the time $t$ during two heating and cooling cycles with $\vartheta$\,=\,10$^{\circ}$C/min for heating and cooling. (b) Reference XRD patterns for all crystalline GeTe phases, tellurium, and germanium.\cite{Shelimova1965,Karbanov1968,Goldak1966,Bradley1924,Cooper1962} The insets show the angular ranges and Miller indices which allow for a distinction between $\alpha$- and $\beta$-GeTe. (c) The \textit{in-situ} XRD pattern of flakes from a crystalline GeTe sputter target shows the transition from $\alpha$- to $\beta$-GeTe during the first heating cycle and back to $\alpha$-GeTe during the following cooling (the relevant temperature ranges are marked by white dashed rectangles). (d) The individual XRD patterns corresponding to the 2$\theta$- and $T$-range of the $\alpha$-to-$\beta$ transition show the peak doublets and singlets with the reflexes marked in green and red, respectively.}
\end{figure}

Fig.~\ref{fig:tempcurve}(a) shows the applied temperature curve. The XRD chamber temperature $T$ is plotted as a function of the time $t$ during two heating and cooling cycles. The samples were heated and cooled at a rate of $\vartheta$\,=\,10$^{\circ}$C/min. At each $T$, the chamber was held for 6\,min total, which includes 1\,min for equilibration and 5\,min for the actual XRD scan. 

In the case of bulk stoichiometric compounds, we expect the crystallization of the initially A-GeTe to the $\alpha$ phase. $\alpha$-GeTe will then remain stable to $T_{C,2}$\,=\,357$^{\circ}$C.\cite{Bletskan2005} Above this temperature, the $\beta$ phase becomes stable. If GeTe is rich in tellurium (>\,50.9\%), A-GeTe crystallizes to the $\gamma$ phase. At elevated temperatures, a $\gamma$-to-$\beta$ transition can be observed.\cite{Bletskan2005} An overview of the crystalline structures and the corresponding reference patterns of GeTe are given in Tab.~\ref{tab:gete} and Fig.~\ref{fig:tempcurve}(b), respectively. Additionally, the reference patterns of crystalline Te and Ge, which are known impurities observed in GeTe,\cite{Caldwell2010,Yarema2018,Jost2015} are shown.

\begin{table}
\caption{\label{tab:gete}Overview of the crystalline phases observed for bulk GeTe. As mentioned in Sec.~\ref{sec:level1intro}, $\alpha$- and $\gamma$-GeTe are stable at room temperature, while $\beta$-GeTe is the high-temperature phase.}
\begin{ruledtabular}
\begin{tabular}{lllll}
Phase & Crystal system & Space group no. & Space group & Ref.\\
\hline
$\alpha$ & trigonal & 160 & R3m & \onlinecite{Goldak1966}\\
$\beta$ & cubic & 225 & Fm$\bar{3}$m & \onlinecite{Shelimova1965}\\
$\gamma$ & orthorhombic & 62 & Pnma & \onlinecite{Karbanov1968}\\
\end{tabular}
\end{ruledtabular}
\end{table}

During the first heating, we collected an XRD pattern every 25$^{\circ}$C for $T$\,$\leq$\,200$^{\circ}$C and every 10$^{\circ}$C for 200$^{\circ}$C\,<\,$T$\,$\leq$\,450$^{\circ}$C. This was based on prior knowledge of the crystallization with $T_{C,1}$\,>\,200$^{\circ}$C observed for small nanoparticles (\textit{cf}.\,Tab.~\ref{tab:temp}) and the $\alpha$-to-$\beta$ transition $T_{C,2}$\,<\,400$^{\circ}$C as reported for bulk GeTe.\cite{Bletskan2005} Since we focused on monitoring the reversible crystalline-to-crystalline transition and no further events were expected for GeTe at lower temperatures during repeated cooling and heating, we adapted our temperature intervals accordingly. Thus, we chose $\Delta T$\,=\,10$^{\circ}$C for 450$^{\circ}$C\,$\geq$\,$T$\,$\geq$\,350$^{\circ}$C and $\Delta T$\,=\,25$^{\circ}$C for $T$\,<\,350$^{\circ}$C.

For a reference value regarding the $\alpha$-to-$\beta$ transition temperature of bulk GeTe, we characterized flakes of a crystalline GeTe sputter target with \textit{in-situ} XRD as described above. The temperature-dependent diffractograms are shown in Fig.~\ref{fig:tempcurve}(c). The transitions from a peak doublet to a single peak for both 2$\theta$\,=\,24\,-\,27$^{\circ}$ and 2$\theta$\,=\,41\,-\,44$^{\circ}$ allow for the confirmation of the $\beta$ phase of GeTe. Based on the XRD scans taken every 10$^{\circ}$C, $T_{C,2}$ is extracted as 380$^{\circ}$C [Fig.~\ref{fig:tempcurve}(d)]. This matches the transition temperature of GeTe with a Te content between 50.2 and 50.5\% given by the phase diagram in Ref.~\onlinecite{Bletskan2005}. Hence, we can conclude that the reference sample is stoichiometric.

\section{\label{sec:level1result}Results and discussion}

In the following, we will discuss the temperature-dependent XRD patterns for three nanoparticle-based samples. First, we will focus on the structural evolution of the GeTe particles from synthesis 1, which showed a broad size distribution. Second, we will analyze the diffractograms obtained for the particles from synthesis 2, which had a narrower size distribution. Also, it provided two samples with sizes smaller and larger than the average size of the particles from synthesis 1.

All diffractograms were normalized by dividing the intensity values by the maximum intensity for each particular diffractogram, meaning $I$/$I_{\textnormal{max}}$, which allows for a better graphical representation. Additionally, the patterns are displayed with a constant offset [($I$/$I_{\textnormal{max}}$)+\,0.25 between each diffractogram] to show the structural evolution of the sample over time. Such waterfall plots facilitate the interpretation of time-dependent XRD data since interpolation as used in 2D contour plots [\textit{cf.} Fig.~\ref{fig:tempcurve}(c)] 
is avoided. Nevertheless, it has to be noted that the time axis has been adapted to follow the XRD patterns. Thus, the spacing is not necessarily equal. This is due to the fact that we chose $\Delta T$\,=\,25$^{\circ}$C for $T$\,$\leq$\,200$^{\circ}$C between each measurement in heating 1 as well as for $T$\,$\leq$\,350$^{\circ}$C in cooling 1, heating 2, and cooling 2. For higher temperatures in each cycle, we chose $\Delta T$\,=\,10$^{\circ}$C for a better resolution. These temperature choices were defined by the expected structural transitions (\textit{cf.} Tab.~\ref{tab:temp}).

In the discussion of all time-dependent XRD diffractograms, we focus on three angular ranges of 2$\theta$ to investigate $T_{C,1}$ and $T_{C,2}$: 
\begin{itemize}[noitemsep]
\item 24-27$^{\circ}$: transition from (003)/(021)-doublet to (111)-singlet marks $\alpha$-to-$\beta$ transition (and \textit{vice versa}),
\item 29-30$^{\circ}$: appearance of (202)-peak marks crystallization, and
\item 41-44$^{\circ}$: transition from (024)/(220)-doublet to (220)-singlet marks $\alpha$-to-$\beta$ transition (and \textit{vice versa}).
\end{itemize}
Since the doublet-to-singlet transition is more pronounced between 41 and 44$^{\circ}$ [\textit{cf.} reference in Fig.~\ref{fig:tempcurve}(b)], we use this angular range to identify $T_{C,2}$. The transition temperatures are marked by dashed rectangles in Figs.~\ref{fig:xrdhanbing}-\ref{fig:xrdy877} and the values for $T_{C,1}$, $T_{C,2}$, and $T_{C,2'}$ are noted next to the pattern. $T_{C,2}$ refers to the $\alpha$-to-$\beta$ transition during heating and $T_{C,2'}$ refers to the reverse transition, $\beta$ to $\alpha$, during cooling.

For orientation, all diffractograms we will discuss below are color-coded with respect to the heating curve and the temperature scaling. In addition to the transition temperatures $T_{C,1}$, $T_{C,2}$, and $T_{C,2'}$, we mark the transition points between each cycle, \textit{i.e.} $T_{\textnormal{min}}$\,=\,75$^{\circ}$C and $T_{\textnormal{max}}$\,=\,450$^{\circ}$C on the right of the XRD patterns. 

\subsection{\label{sec:level2resultHB}\textit{In-situ} XRD on polydisperse GeTe nanoparticles}

\begin{figure}
\includegraphics{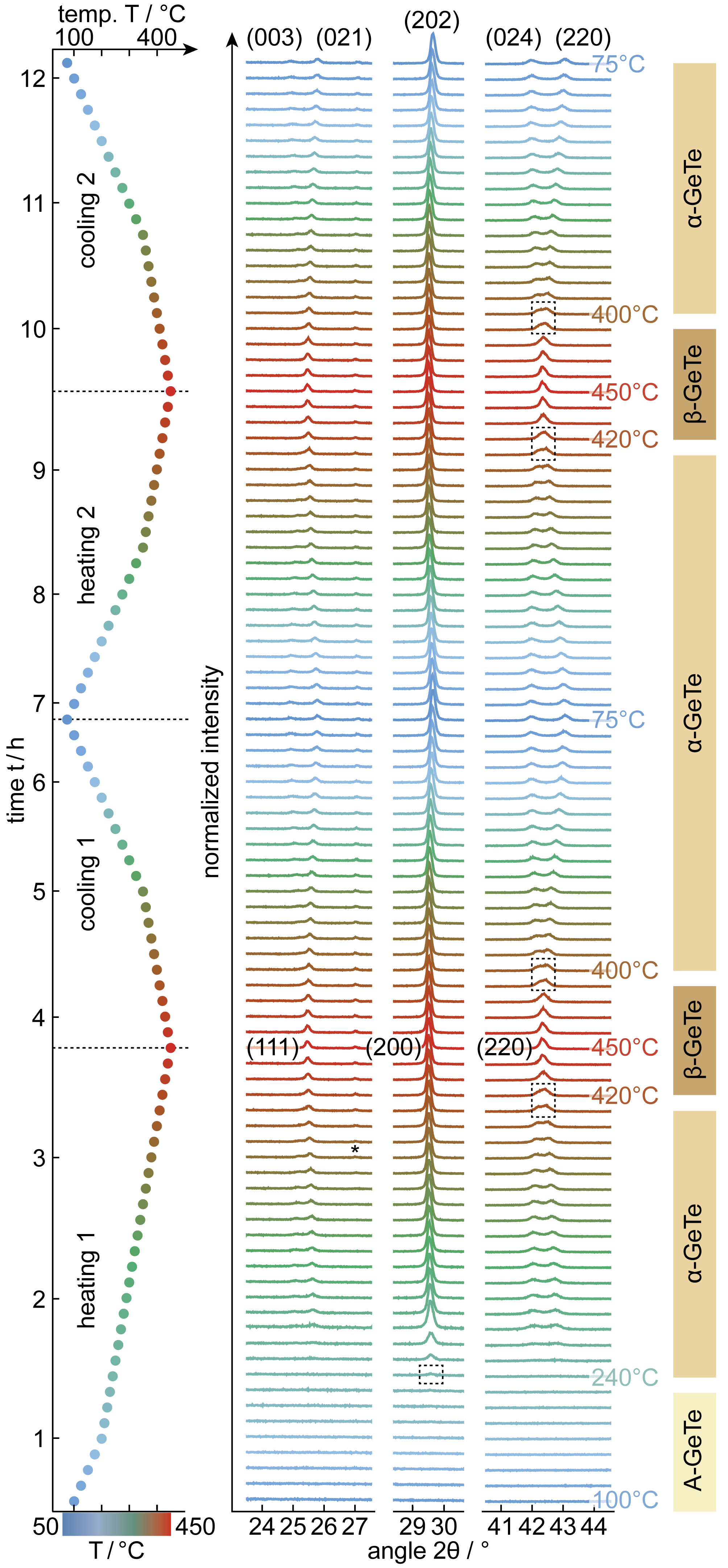}
\caption{\label{fig:xrdhanbing} Structural evolution of GeTe nanoparticles from synthesis 1 with a diameter of 5.5\,$\pm$\,1.6\,nm [green histogram in Fig.~\ref{fig:synthesis}(f)]. Each XRD pattern is normalized to the maximum peak intensity ($I$/$I_{\textnormal{max}}$). The colors represent the scan temperature $T$, as shown on the left side. Upon heating, amorphous GeTe (A-GeTe) crystallizes into $\alpha$-GeTe, $T_{C,1}$\,=\,240$^{\circ}$C, and relaxes into the $\beta$-phase. The transition between $\alpha$- and $\beta$-GeTe continues during further heating and cooling with $T_{C,2}$\,=\,420$^{\circ}$C and $T_{C,2'}$\,=\,400$^{\circ}$C. The small peak at 2$\theta$\,$\approx$\,27$^{\circ}$ (black asterisk) could indicate a small amount of crystalline Te or Ge impurity, \textit{cf.} Fig.~\ref{fig:tempcurve}(b).}
\end{figure}

\begin{figure}
\includegraphics{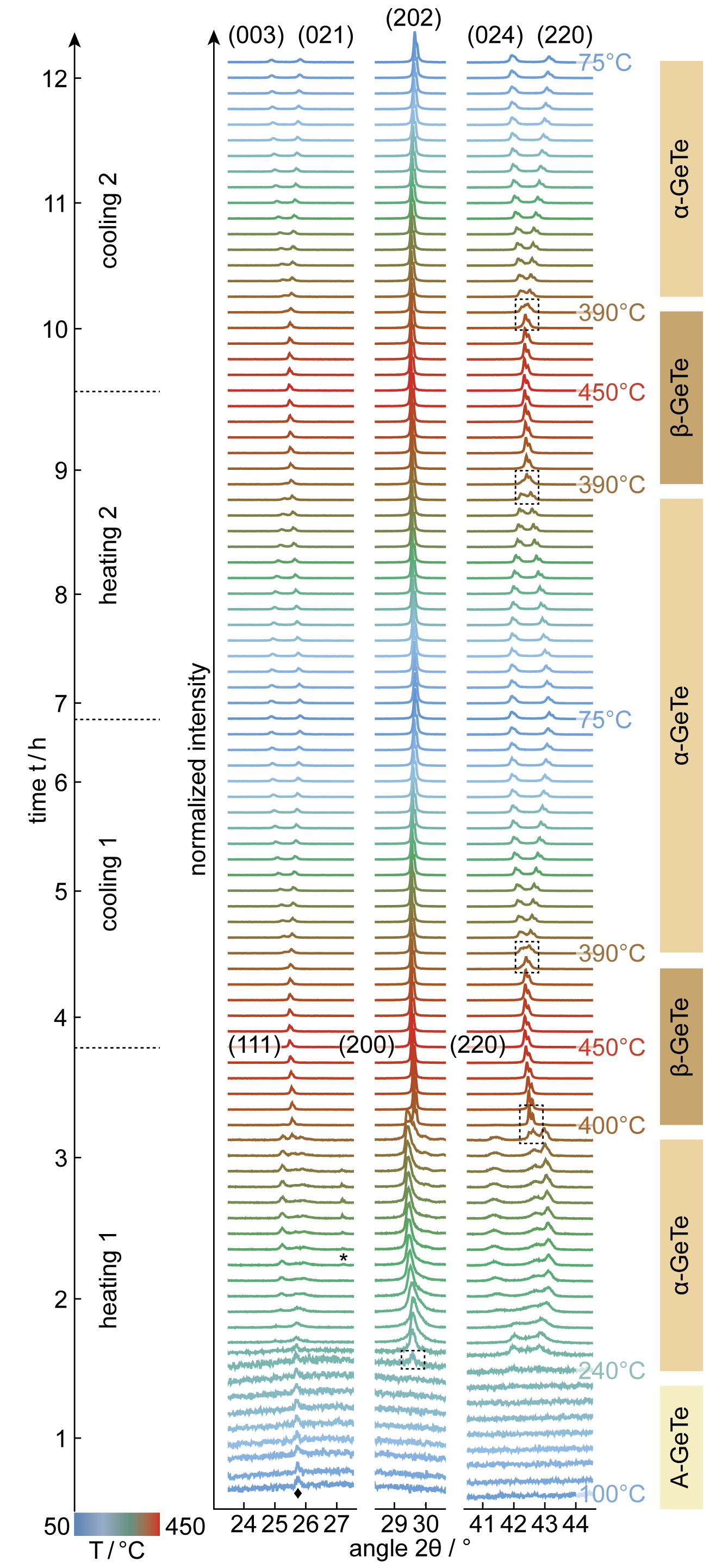}
\caption{\label{fig:xrdy1215} Structural evolution of A-GeTe nanoparticles prepared by synthesis 2 and with a diameter of 4.8\,$\pm$\,0.6\,nm [blue histogram in Fig.~\ref{fig:synthesis}(f)]. Normalization, temperature profile, and color code are similar to Fig.~\ref{fig:xrdhanbing}. Upon heating the A-GeTe nanoparticles, they crystallize into $\alpha$-GeTe at $T_{C,1}$\,=\,240$^{\circ}$C and relax into the $\beta$-phase at $T_{C,2}$\,=\,400$^{\circ}$C. The transition between $\alpha$- and $\beta$-GeTe continues during further heating and cooling with $T_{C,2}$\,=\,$T_{C,2'}$\,=\,390$^{\circ}$C. We ascribe the small peak at 2$\theta$\,$\approx$\,27.2$^{\circ}$ (black asterisk) to a small amount of crystalline Te or Ge impurity, \textit{cf.} Fig.~\ref{fig:tempcurve}(b). The small peak at 2$\theta$\,$\approx$\,25.7$^{\circ}$ (black rhombus) could not be matched with any reference diffractograms given (\textit{cf.} Appendix~\ref{sec:App2}). However, carbon shows a strong peak at this angle and could have contaminated the sample.}
\end{figure}

The diffractograms of the GeTe nanoparticles from synthesis 1, which led to a broad size distribution [\textit{cf.} Fig.~\ref{fig:synthesis}(f)], are shown in Fig.~\ref{fig:xrdhanbing}. The XRD patterns start at $T$\,=\,100$^{\circ}$C in the first heating cycle. 
During heating 1, a narrowing of the intensity peak close to 2$\theta$\,=\,30$^{\circ}$ as well as a convergence of the (024)/(220)-doublet can be seen. First, we focus on the interpretation of the width of the (202) and (200)-peak of the $\alpha$ and $\beta$ phase, respectively. In a general and simplified consideration, the full width at half maximum (FWHM or $w$) of an XRD peak can be related to the lattice strain and the crystallite size $D$. It has to be noted, that $D$ is not necessarily identical with the particle size $d$. In case of coalescence for example, $D$ can be larger than the initial $d$. Hence, without a high-resolution TEM investigation it is difficult to judge whether the particle or domain sizes are determined via XRD. This ambiguity has to be kept in mind when the term \textit{crystallite} is used for $D$.\cite{Girgsdies2015} Nevertheless, its size can theoretically be estimated from the well-known Scherrer equation: 
\[D = [K \cdot \lambda]/[w \cdot cos(\theta)]\]

with $K$ being the shape factor or Scherrer constant (often approximated by 0.9), $\lambda$ being the X-ray wavelength in nm, and $\theta$ being the angle of diffraction in rad.\cite{Girgsdies2015} The observed peak width $w_{\textnormal{obs}}$ in rad has to be corrected by subtracting the instrumental broadening $w_{\textnormal{instr}}$ in rad:
\[w = w_{\textnormal{obs}} - w_{\textnormal{instr}}.\]

In our case, $w_{\textnormal{instr}}$\,$\approx$\,0.1$^{\circ}$ 
and thus, up to 72\% of $w_{\textnormal{obs}}$. Therefore, it would be necessary to collect the diffractograms with a much higher resolution (\textit{e.g.} at a synchrotron) to obtain a better estimate of $D$. Nevertheless, a qualitative approximation of $D$ based on the XRD patterns is possible. In a simplified picture, the initial particle diameter $d$\,$\approx$\,5.5\,nm determined by TEM would imply $w_{\textnormal{obs}}$\,$\approx$\,1.6$^{\circ}$ 
(assuming $D$\,=\,$d$, $K$\,=\,0.9, $w_{\textnormal{instr}}$\,=\,0.1$^{\circ}$, and $\lambda$\,=\,0.15406\,nm). In contrast, Fig.~\ref{fig:xrdhanbing} shows 0.14$^{\circ}$\,$\leq$\,$w_{\textnormal{obs}}$\,$\leq$\,0.2$^{\circ}$. Thus, it seems very likely that the GeTe nanoparticles coalesced during the initial crystallization. Thereby, it has to be emphasized that XRD patterns represent only an averaged signal, and no conclusion on the individual particles can be drawn. 

Between $T$\,=\,240 and 300$^{\circ}$C the peak width decreases further, indicating continued growth of $D$. This could relate to further coalescence or growth of larger grains at the expense of smaller ones.\cite{Polking2011,Yarema2018} Further thermal treatment showed no influence on the width of the (202)- or (200)-peak of $\alpha$- or $\beta$-GeTe, respectively. Thus, it can be assumed that grain growth stopped. Apart from the narrowing of the (202)-peaks, a slight shift to smaller angles upon heating can be seen. During cooling, the peak position shifts back to its initial value. A similar behavior can be seen for the second heating and cooling. This can be rationalized by the relaxation of the distorted $\alpha$-phase to the cubic $\beta$ phase.\cite{Polking2011,Boschker2017} 

For $T$\,=\,260$^{\circ}$C, the (024)/(220)-doublet becomes clearly visible and converges smoothly towards the (220)-singlet for increasing $T$\,$\geq$\,360$^{\circ}$C. This indicates a slow transition from the rhombohedrally distorted $\alpha$-phase to the relaxed cubic $\beta$-phase. Something similar has been observed for larger GeTe nanoparticles, which were crystalline after synthesis.\cite{Polking2011} During the first cooling cycle, the gradual splitting of the (220)-peak towards the (024)/(220)-doublet can be seen. The related $\beta$-to-$\alpha$-transition temperature was $T_{C,2'}$\,=\,400$^{\circ}$C according to the diffractogram. A similar behavior can be found for heating and cooling 2, where the (024)/(220)-doublet transitions into the (220)-singlet at $T_{C,1}$\,=\,420$^{\circ}$C and the high-temperature phase transitions back into the room-temperature phase at $T_{C,2'}$\,=\,400$^{\circ}$C. Thus, the transition between the two C-GeTe phases is reversible for the sample based on nanoparticles from synthesis 1.

The aforementioned reversible change is less evident for the (003)/(021)-doublet to (111)-singlet transition at 2$\theta$\,=\,24$\ldots$27$^{\circ}$. This was expected due to the low intensity of these peaks in the reference [\textit{cf.} Fig.~\ref{fig:tempcurve}(b)]. Nevertheless, the transition is visible, but does not allow for a determination of the transition temperatures $T_{C,2}$ and $T_{C,2'}$.

\subsection{\label{sec:level2resultMY}\textit{In-situ} XRD on nanoparticles with narrower size distribution}

\begin{figure}
\includegraphics{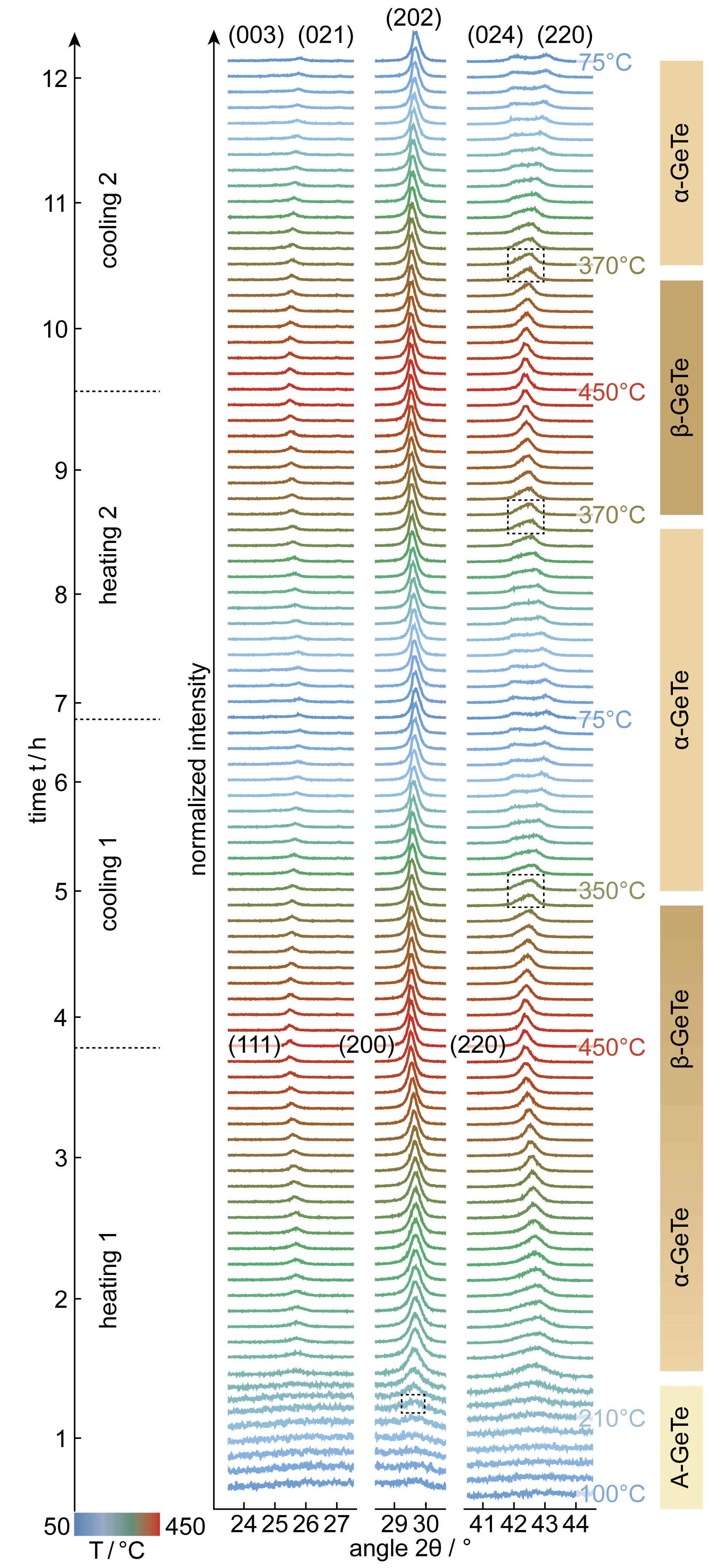}
\caption{\label{fig:xrdy877} Structural evolution of A-GeTe nanoparticles from synthesis 2 with a diameter of 6.9\,$\pm$\,0.9\,nm [red histogram in Fig.~\ref{fig:synthesis}(f)]. Normalization, temperature profile, and color code are similar to Figs.~\ref{fig:xrdhanbing} and \ref{fig:xrdy1215}. Upon heating, A-GeTe crystallizes at $T_C$\,=\,210$^{\circ}$C. Presumably, $\alpha$-GeTe relaxes into $\beta$-GeTe since the transition back to the $\alpha$ phase can be found at $T_{C,2'}$\,=\,350$^{\circ}$C. During further heating and cooling the crystalline-to-crystalline transitions can be found at $T_{C,2}$\,=\,$T_{C,2'}$\,=\,370$^{\circ}$C.}
\end{figure}

In Figs.~\ref{fig:xrdy1215} and \ref{fig:xrdy877} the diffractograms of the nanoparticles from synthesis 2 are shown. These two samples had a smaller and larger average size and a narrower size distribution than the GeTe nanoparticles discussed in the previous section. 

While the XRD patterns were collected, normalized, and plotted similarly to Fig.~\ref{fig:xrdhanbing}, it is obvious that the diffractograms in Figs.~\ref{fig:xrdy1215} and \ref{fig:xrdy877} for the amorphous samples with 100$^{\circ}$C\,$\leq$\,$T$\,$\leq$\,250$^{\circ}$C are noisier. One possible reason could be that the total amount of sample from synthesis 2 was lower compared to samples from synthesis 1 due to different sample contributions from the TOP and oleic acid ligands (\textit{cf.} Section \ref{sec:level2Synth}). Another reason could be the normalization $I$/$I_{\textnormal{max}}$. This leads to a pronounced increase of the signal-to-noise ratio (SNR) for XRD patterns above $T_{C,1}$, where $I_{\textnormal{max}}$\,$\gg$\,$I$, but does not affect the SNR of amorphous diffractograms (\textit{cf.} Appendix~\ref{sec:App1}).

In Fig.~\ref{fig:xrdy1215}, the first sign of crystallization into $\alpha$-GeTe can be observed for $T$\,=\,240$^{\circ}$C which matches what we already observed in Fig.~\ref{fig:xrdhanbing}. Again, the peak width indicates a coalescence of particles prior to or during crystallization. Already the XRD pattern for the next temperature step shows a clear (024)/(220)-doublet. Further heating leads to a quick divergence of the two peaks, while the high-angle peak seems to overlap with another small peak. At the same time, the (202) peak shifts quite strongly to lower angles and a weak (003)/(021)-doublet becomes visible. The latter shows a prompt divergence and an overlapping additional peak (for $T$\,=\,370\,-\,390$^{\circ}$C) as well. The described trends remain for all three diffractive signatures of the $\alpha$-phase until $T$\,=\,400$^{\circ}$C. For this temperature, a prompt shift of the central peak occurs and both doublet-to-singlet transitions are observed. Thus, we define the onset of the $\alpha$-to-$\beta$-transition at $T_{C,2}$\,=\,400$^{\circ}$C. Further heating leads to a relaxation of the peaks and the subsequent cooling shows the smooth and reversible transition between $\beta$- and $\alpha$-GeTe. This behavior is similar to what was seen in Fig.~\ref{fig:xrdhanbing}. Only the transition temperatures are slightly lower with $T_{C,2'}$\,=\,390$^{\circ}$C and  $T_{C,2}$\,=\, $T_{C,2'}$ from there on.

It has to be noted, that the (024)/(220)-doublet shows very small split peaks. This could be related to the fact that we used CuK$_{\alpha,1}$ and CuK$_{\alpha,2}$. Since the peaks are sharper compared to what we found for the diffractograms in Fig.~\ref{fig:xrdhanbing}, this effect might appear more strongly for the sample investigated here. Furthermore, the narrower peaks shown in Fig.~\ref{fig:xrdy1215} indicate a larger crystallite size $D$ and thus, more pronounced coalescence than observed for the GeTe particles from synthesis 1 can be assumed. This is surprising since oleic acid ligands are used for the particles from synthesis 2. These molecules are much longer than the TOP ligands, which are shown in Fig.~\ref{fig:synthesis}(c), and thus, a more pronounced nanoparticle separation and potentially less coalescence could be expected.\cite{Yarema2018}

Similar to the diffractograms in Fig.~\ref{fig:xrdhanbing}, a small additional peak not matching the $\alpha$- or $\beta$-lattice was found. It appears at 2$\theta$\,$\approx$\,27.2$^{\circ}$ for $T$\,$\geq$\,310$^{\circ}$C, but it disappears for $T$\,$\geq$\,380$^{\circ}$C and does not reoccur throughout the following temperature treatment. Due to the angular position of this peak, it could indicate traces of crystalline Te or Ge in the sample. Since the most pronounced peaks for both crystalline patterns are almost overlapping [\textit{cf.} Fig.~\ref{fig:tempcurve}(b)], an unambiguous identification is not possible. Nevertheless, Te impurities have been reported for annealed initially amorphous GeTe nanoparticles.\cite{Caldwell2010} Additionally, segregated Te has been observed as a result of surface oxidation of GeTe (\textit{cf.} Appendix~\ref{sec:App2}).\cite{Kolb2019} 

In Fig.~\ref{fig:xrdy877}, a very weak crystalline signal can already be identified for $T$\,=\,100$^{\circ}$C. The (202)-peak is very broad and flat, thus indicating a much smaller $D$ than what we found for the smaller GeTe nanoparticles (\textit{cf.} Figs.~\ref{fig:xrdhanbing} and \ref{fig:xrdy1215}). Due to this initial (partial) crystallization of the particles, the determination of $T_{C,1}$ is difficult. Therefore, we use the enhancement of this hump at 2$\theta$\,$\approx$\,29.6$^{\circ}$ and the simultaneous onset of a crystalline signal in the high-angle range, which we define for $T_{C,1}$\,=\,210$^{\circ}$C. However, this is less reliable than $T_{C,1}$ defined for the samples based on the smaller GeTe nanoparticles discussed above.

Further heating leads to more pronounced peaks for the three considered angular ranges. Thereby, the (202)-peaks narrow, while the peak width $w_{\textnormal{obs}}$ remains relatively broad compared to what is shown in Figs.~\ref{fig:xrdhanbing} and \ref{fig:xrdy1215}. This would imply that coalescence did not progress as far as for these samples. If we could apply the simplified logic of longer ligands resulting in a lower degree of particle sintering, we would expect $D_{\textnormal{5.5}}$\,>\,$D_{\textnormal{6.9}}$ (particles with an average diameter $d$\,$\approx$\,5.5\,nm and TOP ligands \textit{versus} particles with $d$\,$\approx$\,6.9\,nm and oleate ligands).

Similar to what we observed before, the (202)- and (200)-peak, respectively, shift smoothly. However, during the first heating, we cannot determine $T_{C,2}$. Instead, during cooling a broadening and subsequent peak splitting from the (220)-singlet to the (024)/(220)-doublet can be observed, starting at $T_{C,2'}$\,=\,350$^{\circ}$C. The following $\alpha$-to-$\beta$-transition sets in at $T_{C,2'}$\,=\,370$^{\circ}$C with the transition back to $\alpha$ at the same temperature during the second heating cycle. The (003)/(021)-doublet cannot be observed before the end of the last cooling, \textit{i.e.} $T$\,=\,75$^{\circ}$C. 

\section{\label{sec:level1outlook}Conclusion}
We synthesized ultrasmall nanoparticles of the phase-change material GeTe with diameters below 10\,nm. In the literature, these sizes have been identified to show size-dependent material properties. We studied the crystallization behavior of drop-casted nanoparticle films using \textit{in-situ} XRD while heating the films under a nitrogen atmosphere. All nanoparticle-based samples showed crystallization to $\alpha$-GeTe followed by a crystalline-to-crystalline transition to the high-temperature $\beta$-phase of GeTe. 
During cooling, this transition was reversible and could be repeated for a second heating and cooling cycle. All samples showed increased crystallization temperatures $T_{C,1}$ and $T_{C,2}$ compared to bulk GeTe.

\begin{figure}
\includegraphics{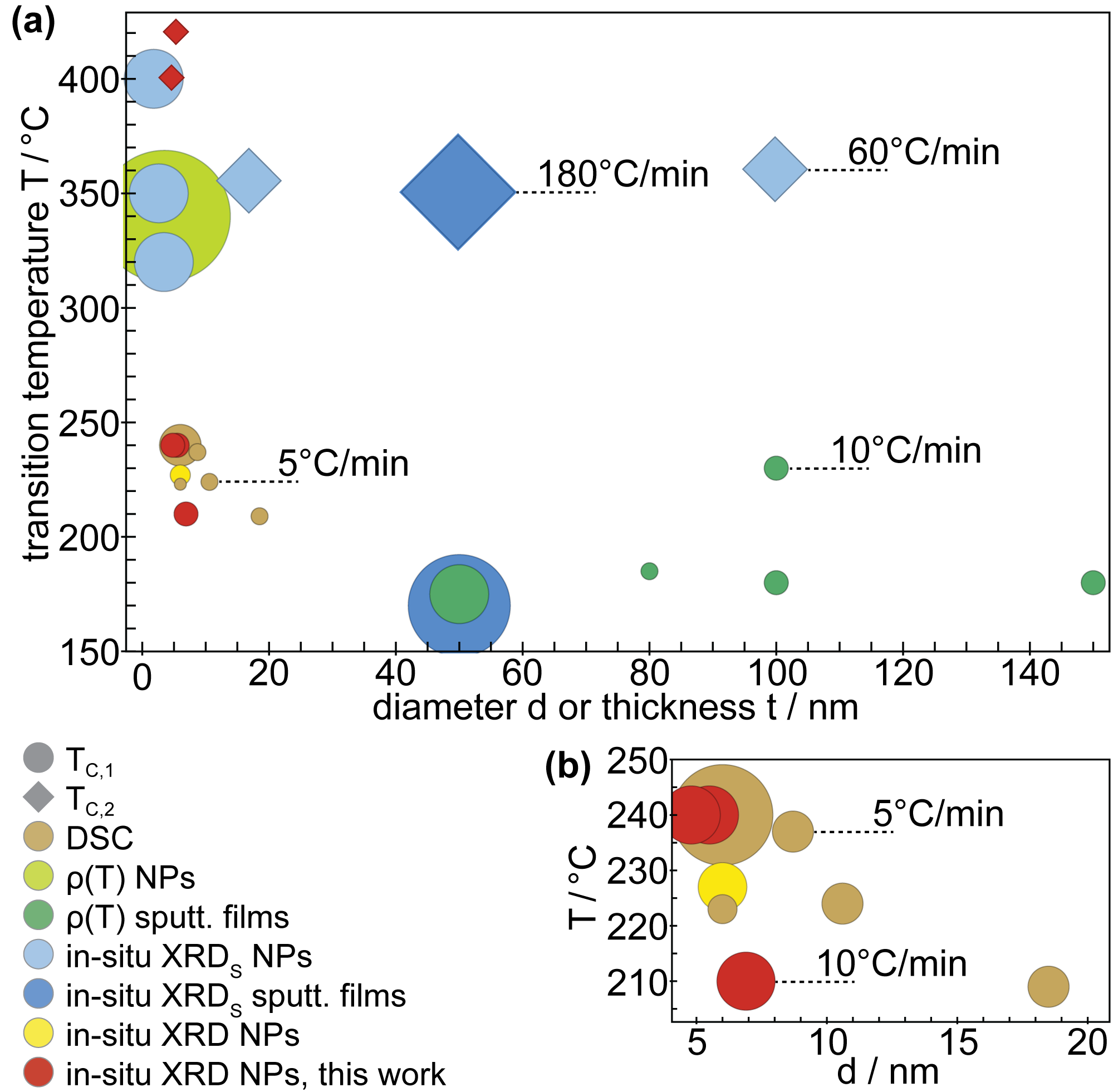}
\caption{\label{fig:Toverview} (a) Comparison of the GeTe transition temperatures $T_{C,1}$ (disk) and $T_{C,2}$ (diamonds) determined for different particle sizes $d$ and film thicknesses $t$. The colors represent the different characterization techniques and samples. The larger the symbol area, the higher was the applied heating rate $\vartheta$. Exemplary values of $\vartheta$ are noted in the plot. (b) A magnified plot for the section in (a) where most $T_{C,1}$ were determined for small $d$.}
\end{figure}

Fig.~\ref{fig:Toverview} compares our results to the values obtained by previous studies (\textit{cf.} Tab.~\ref{tab:temp}). While we observed coalescence, similar to the literature, we list the transition temperatures as a function of the initial particle size $d$. Fig.~\ref{fig:Toverview} reveals that the literature values follow a general trend of an increasing $T_{C,1}$ and $T_{C,2}$ for decreasing $d$ below 10\,nm. However, quantification of the size-dependent $T_{C,1}$ and $T_{C,2}$ is complicated by coalescence. We propose to perform a similar study with separated nanoparticles, \textit{e.g.} by using atomic layer deposition. Furthermore, \textit{in-situ} TEM, \textit{in-situ} Raman spectroscopy, and ultrafast DSC could give further insights into the crystallization of individual nanoparticles.\cite{Chen2016,Polking2011,Pries2019} This would be especially interesting for GeTe, since a decrease of $T_{C,1}$ has been found for decreasing particle size $d$ for Ge$_2$Sb$_2$Te$_5$, which is one of the most prominent phase-change materials.\cite{Shportko2009,Wuttig2017} In contrast, Ref.~\onlinecite{Raoux2008} reported an increased $T_{C,1}$ for sputter-deposited doped GeSb, Sb$_2$Te, and Ge$_2$Sb$_2$Te$_5$ films with thicknesses $t$\,<\,10\,nm. Nevertheless, Ref.~\onlinecite{Simpson2010} showed for Ge$_2$Sb$_2$Te$_5$ that this behavior can be ascribed to (capping-dependent) strain in the thin films. The crystallization of nanoparticles will likely be influenced by the large surface-to-volume ratio and presumably, significant strain. Moreover, Kolb \textit{et al.} reported on the nucleation of non-oxidized bulk GeTe at 230$^{\circ}$C. In this context, the crystallization at 180$^{\circ}$C was found to be induced by surface oxidation and related elemental segregation, which led to Te serving as nucleation sites.\cite{Kolb2019}

Apart from the initial crystallization, our study focused on the reversible $\alpha$-to-$\beta$-transition which we observed for the three nanoparticle samples. These particles were initially amorphous, synthesized following two different protocols, and had different diameters. The transition temperatures were higher compared to the thin films and crystalline nanoparticles discussed in the literature (\textit{cf.} Tab.~\ref{tab:temp}), even if particle coalescence is assumed. It is promising that samples based on very small solution-deposited amorphous nanoparticles still show reversible phase change behavior along with an increased bandgap and tunable refractive index.\cite{Michel2020} This is potentially useful for the scalability of active photonic, phase-change random access memory, and optical data storage.

\begin{acknowledgments}
This project was funded by the European Research Council under the European Union's Seventh Framework Program (FP/2007-2013)/ERC Grant Agreement Number 339905 (QuaDoPS Advanced Grant). A.-K.U.M. acknowledges funding from the ETH Zurich Postdoctoral Fellowship Program and the Marie Curie Actions for People COFUND Program (Grant 17-1 FEL-51). M.Y. acknowledges funding from the SNF Ambizione Fellowship (No.\,161249). The authors thank H.~Rojo Sanz, S.~Meyer, and I.~Giannopoulos for technical assistance and P.~Knüsel, M.D.~Wörle, and J.~Schawe for fruitful discussions. TEM measurements were performed at the Scientific Center for Optical and Electron Microscopy (ScopeM) at ETH Zurich.
\end{acknowledgments}

\section*{Data Availability}
The data that support the findings of this study are available from the corresponding author upon reasonable request.

\appendix

\section{\label{sec:App1}Full-range XRD diffractograms}
\begin{figure}
\includegraphics{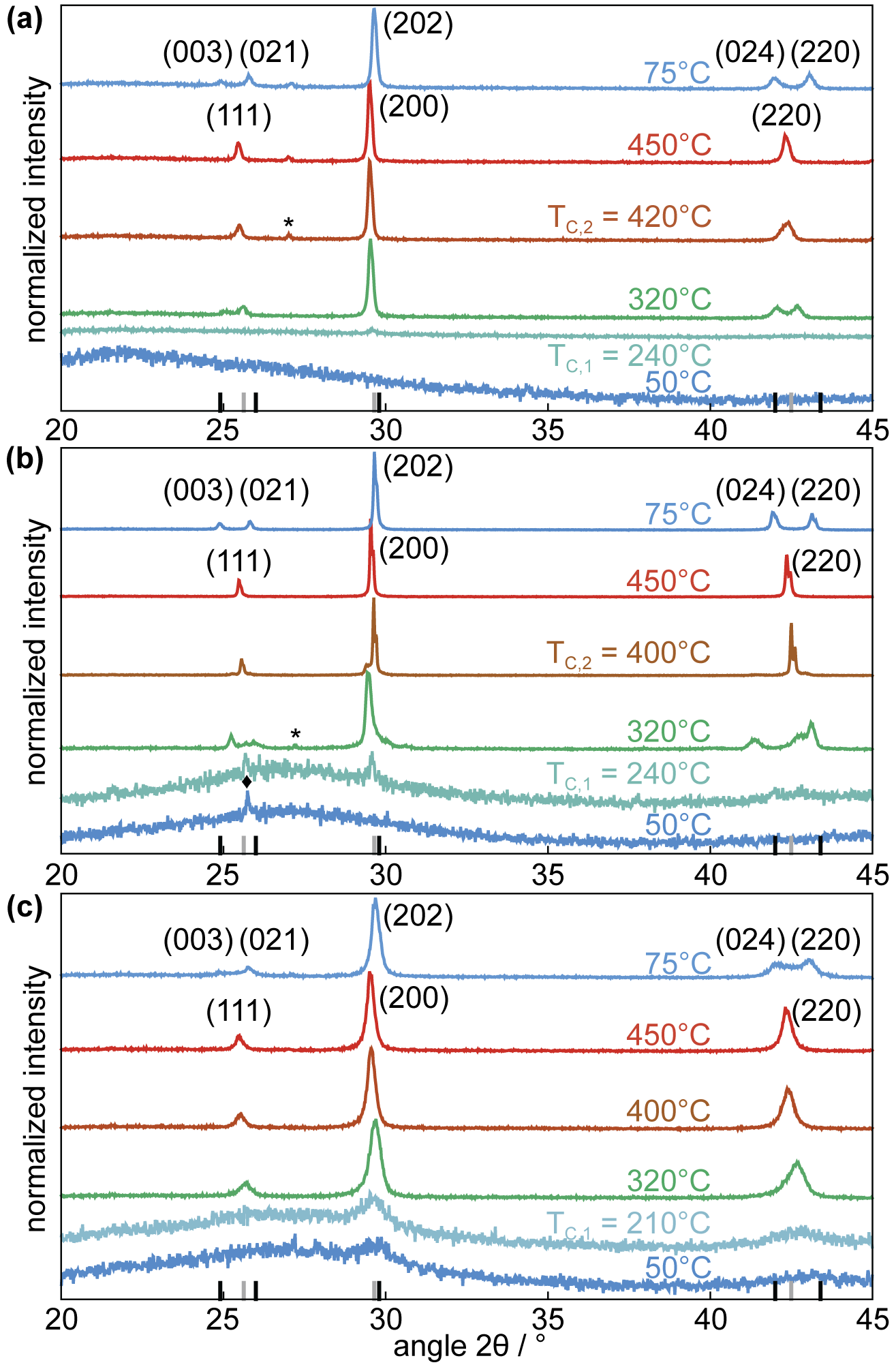}
\caption{\label{fig:xrdappendix} Diffractograms of A-GeTe nanoparticles shown in (a), Fig.~\ref{fig:xrdhanbing}, (b), Fig.~\ref{fig:xrdy1215}, and (c), Fig.~\ref{fig:xrdy877}, for the full angular range of detection. The normalization and color code are similar to Figs.~\ref{fig:xrdhanbing}\,-\,\ref{fig:xrdy877}. The Miller indices are noted next to the peaks. The reference patterns for $\alpha$- and $\beta$-GeTe are marked in black and gray, respectively, at the bottom of each subfigure. Additional features are marked similarly to Figs.~\ref{fig:xrdhanbing} and \ref{fig:xrdy1215}: (a), (b) black asterisk for Te or Ge impurities, and (b) black rhombus for an unidentified small peak, potentially associated with carbon.}
\end{figure}

For comparison, the full XRD patterns are shown for examplary temperatures in Fig.~\ref{fig:xrdappendix} with Fig.~\ref{fig:xrdappendix}(a) related to Fig.~\ref{fig:xrdhanbing}, Fig.~\ref{fig:xrdappendix}(b) related to Fig.~\ref{fig:xrdy1215}, and Fig.~\ref{fig:xrdappendix}(c) related to Fig.~\ref{fig:xrdy877}. The XRD diffractograms are shown for: 
\begin{itemize}[noitemsep]
\item 50$^{\circ}$C, which is the lowest $T$ for which XRD was conducted,
\item the respective $T_{C,1}$,
\item 320$^{\circ}$C for $\alpha$-GeTe related to heating 1,
\item the respective $T_{C,2}$, if applicable,
\item 450$^{\circ}$C, which is the highest $T$ for which XRD was conducted, and
\item 75$^{\circ}$C for $\alpha$-GeTe after cooling 1.
\end{itemize}

\section{\label{sec:App2}Identification of XRD peak at 2$\theta$\,$\approx$\,25.7$^{\circ}$}
Considering the particles right after synthesis, the XRD patterns in Fig.~\ref{fig:xrdy1215} show a small peak at 2$\theta$\,$\approx$\,25.7$^{\circ}$. According to the potential reference patterns, this would only match $\beta$-GeTe which can be excluded as this phase is expected only at higher temperatures. Furthermore, it has been reported that oxidation and segregation can be expected for annealed GeTe.\cite{Kolb2019} To prevent this, synthesis and XRD measurements were performed oxygen-free. Nevertheless, if oxygen had been present (at elevated temperatures) only amorphous GeO$_x$ and crystalline Te would be expected.\cite{Kolb2019} The small initial XRD peak at about 25.7$^{\circ}$ cannot be explained by this. Another possibility could be residual germanium diiodide which is used as precursor in the synthesis. However, the strongest XRD peaks would be expected at 2$\theta$\,$\approx$\,30.4 and 26.2$^{\circ}$ according to Ref.~\onlinecite{Avilov1968}.

\bibliography{michel_time-dependent-xray-gete-nps_references}

\end{document}